\def\prd{Phys. Rev. D}

\def\apj{Astrophys. J.}

\def\apjs{Astrophys.J.Suppl.}

\documentclass[nofootinbib,floatfix,twocolumn,showpacs,amssymb,amsmath,prd]{revtex4-1} 
\usepackage{hyperref}
\usepackage{breakurl}
\usepackage{graphicx}
\usepackage{color}
\usepackage{bm}
\begin{document}
\title{Limits on anisotropic inflation from the \textit{Planck} data}\thanks{Based on observations obtained with \textit{Planck} (http://www.esa.int/Planck), an ESA science mission with instruments and contributions directly funded by ESA Member States, NASA, and Canada.}
\author{Jaiseung Kim}
\email{kim@mpa-garching.mpg.de}
\affiliation{Max-Planck-Institut f\"{u}r Astrophysik, Karl-Schwarzschild Str. 1, 85741 Garching, Germany}
\affiliation{Niels Bohr Institute, Blegdamsvej 17, DK-2100 Copenhagen, Denmark}
\author{Eiichiro Komatsu}
\affiliation{Max-Planck-Institut f\"{u}r Astrophysik, Karl-Schwarzschild Str. 1, 85741 Garching, Germany}
\affiliation{Kavli Institute for the Physics and Mathematics of the Universe, Todai Institutes for Advanced Study, the University of Tokyo, Kashiwa, Japan 277-8583 (Kavli IPMU, WPI)}
\date{\today}

\begin{abstract}
 Temperature anisotropy of the cosmic microwave background offers a test
 of the fundamental symmetry of spacetime during cosmic
 inflation. Violation of rotational symmetry yields a distinct signature
 in the power spectrum of primordial fluctuations as $P({\mathbf
 k})=P_0(k)[1+g_*(\hat{\mathbf k}\cdot\hat{\mathbf E}_{\rm cl})^2]$,
 where $\hat{\mathbf E}_{\rm cl}$ is a preferred direction in space and
 $g_*$ is an amplitude. Using the \textit{Planck} 2013 temperature maps,
 we find no evidence for violation of rotational symmetry,
 $g_*=0.002\pm 0.016$ (68\%~CL), once the known effects of
 asymmetry of the \textit{Planck} beams and Galactic foreground emission
 are removed.
\end{abstract}
\pacs{98.70.Vc, 98.80.Cq, 98.80.-k}
\maketitle 

Cosmic inflation
\cite{Starobinsky:1980,Sato:1980,Guth:1980,Linde:1981,Albrecht:1982},
an indispensable building-block of the standard model of the universe,
is described by {\it nearly} de Sitter spacetime. The metric charted by
flat coordinates is given by
$ds^2=-dt^2+e^{2Ht}d{\mathbf x}^2$, where $H$ is the expansion rate of
the universe during inflation. This spacetime admits ten isometries:
three spatial translations; three spatial rotations; one time
translation accompanied by spatial dilation ($t\to t-\lambda/H$ and
${\mathbf x}\to e^\lambda{\mathbf x}$ with a constant $\lambda$); and
three additional isometries which reduce to special conformal
transformations in $t\to \infty$. The necessary time-dependence of the
expansion rate, $Ht\to \int H(t')dt'$, breaks the time
translation symmetry hence the spatial dilation symmetry, yielding the
two-point correlation function of primordial fluctuations that is
nearly, but not exactly, invariant under ${\mathbf x}\to
e^\lambda{\mathbf x}$ \cite{Mukhanov:1981}. The magnitude of the
deviation from dilation invariance is limited by that of the
time-dependence of $H$, i.e., $-\dot H/H^2={\cal O}(10^{-2})$. 

In the usual model of inflation, six out of ten isometries remain
unbroken: translations and rotations. Why must they remain unbroken
while the others are broken?  In this paper, we shall test rotational
symmetry during inflation, using the two-point correlation function of
primordial perturbations to spatial curvature, $\zeta$, generated during
inflation. This is defined as a perturbation to the exponent in the
spatial metric, $\int H(t')dt'\to \int H(t')dt'+\zeta({\bf x},t)$. In
Fourier space, we write the two-point function as $\langle\zeta_{\bf
k}\zeta^*_{{\bf k}'}\rangle=(2\pi)^3\delta^{(3)}({\bf k}-{\bf k}')P({\bf
k})$, and $P({\bf k})$ is the power spectrum. Translation invariance,
which is kept in this paper, gives the delta function, while
rotation invariance, which is {\it not} kept, would give $P({\bf
k})\to P(k)$ with $k\equiv |{\bf k}|$. Dilation invariance would give
$k^3P(k)={\rm const}.$, whereas a small deviation, $k^3P(k)\propto
k^{-0.04}$, has been detected from the CMB data with more than
5-$\sigma$ significance \cite{WMAP9:cosmology,Planck_cosmology}.

Following Ref.~\cite{Ackerman:2007}, we write the power spectrum as
$P({\mathbf k})=P_0(k)\left [1+g_{*}(k)\,(\hat {\mathbf k}\cdot \hat
{\mathbf E}_{\mathrm{cl}})^2 \right]$, where $\hat {\mathbf
E}_{\mathrm{cl}}$ is a preferred direction in space, $g_*$ is a
parameter characterizing the amplitude of violation of rotational
symmetry, and $P_0(k)$ is an isotropic power spectrum which depends only
on the magnitude of the wavenumber, $k$. This form is generic, as
it is the leading-order anisotropic correction that remains invariant
under parity flip, ${\bf k}\to -{\bf 
k}$. ``Anisotropic inflation'' models, in which a scalar field is
coupled to a vector field (see
Ref.~\cite{Soda:2012,Maleknejad:2012,Dimastrogiovanni:2010} and
references therein) can produce this form.\footnote{Anisotropic
inflation models produce three-point functions of $\zeta$ which also depend on
$g_*$ \cite{Barnaby:2012,Bartolo_F2:2013,Shiraishi:2013}. The Planck team
uses this property to put model-dependent constraints on $g_*$ from
non-detection of primordial three-point functions \cite{planck2013fnl}.} 
A very long-wavelength perturbation on super-horizon scales can also
produce this form via a three-point function \cite{Schmidt:2012}. 
A pre-inflationary universe was probably chaotic and highly
anisotropic, and thus a remnant of the pre-inflationary anisotropy may
still be detectable \cite{Bartolo:2013}.

We shall ignore a potential $k$ dependence of $g_*$ in this paper. 
We expand $g_*\,(\hat {\mathbf
k}\cdot \hat {\mathbf E}_{\mathrm{cl}})^2$ using spherical harmonics:
\begin{eqnarray} 
g_*\,(\hat {\mathbf k}\cdot \hat{\mathbf E}_{\mathrm{cl}})^2&=&\frac{g_*}{3}+\frac{8\pi}{15}g_*\sum_{M} Y_{2M}^*(\hat{\mathbf E}_{\mathrm{cl}})\,Y_{2M}(\hat{\mathbf k}).
\end{eqnarray}
We then write the power spectrum as
\begin{eqnarray} 
P({\mathbf k})&=&\tilde{P}_0(k)\left [1+\sum_{M}g_{2M} Y_{2M}(\hat{\mathbf k})\right],\label{prim_power_glm}
\end{eqnarray}
where we have absorbed $g_*/3$ into the normalization of the isotropic part,
 $\tilde{P}_0(k)\equiv P_0(k)(1+g_*/3)$, and defined
$g_{2M}\equiv \frac{8\pi}{15}\frac{g_*}{1+g_*/3}Y_{2M}^*(\hat{\mathbf
 E}_{\mathrm{cl}})$ with $g_{2M}$ for $M<0$ given by $g_{2,-M}=(-1)^M\,g^*_{2,M}$. 

There are 5 parameters to be determined from the data. We denote the
parameter vector as ${\bf h}\equiv
\{g_{20},\mathrm{Re}[g_{21}],\mathrm{Im}[g_{21}],\mathrm{Re}[g_{22}],\mathrm{Im}[g_{22}]\}$. 
We search for $\bf h$ in the covariance matrix of the
spherical harmonics coefficients of CMB temperature maps,
$\mathcal C_{l_1m_1,l_2m_2}\equiv \langle a_{l_1m_1}
a^*_{l_2m_2}\rangle$, where $a_{lm}=\int d^2\hat{\bf n}~T(\hat{\bf
n})Y^*_{lm}(\hat{\bf n})$. 
The anisotropic power spectrum of Eq.~\ref{prim_power_glm} gives
\cite{gstar_forecast}
\begin{eqnarray} 
\mathcal C_{l_1m_1,l_2m_2} 
&=&\delta_{l_1l_2}\delta_{m_1m_2}\,C_{l_1}+\imath^{l_1-l_2} (-1)^{m_1} D_{l_1l_2} \nonumber\\
&&\times\sum_{M} g_{2M}\left[\frac{5(2l_1+1)(2l_2+1)}{2\pi}\right]^{\frac{1}{2}}\nonumber\\
&&\times\left(\begin{array}{ccc}2&l_1&l_2\\0&0&0\end{array}\right)\left(\begin{array}{ccc}2&l_1&l_2\\M&-m_1&m_2\end{array}\right),
\end{eqnarray}
where the matrices denote the Wigner 3-$j$ symbols, and
$D_{l_1l_2}\equiv \frac{2}{\pi}\int k^2 dk\,\tilde{P}_0(k)\,g_{Tl_1}(k)\,g_{Tl_2}(k)$ with
$g_{Tl}(k)$ the temperature radiation transfer function. 

In the limit of weak anisotropy, the likelihood of the CMB data given a
model may be expanded as
\begin{eqnarray}
\mathcal{L}&=&\mathcal{L}|_{h=0}+\sum_i\left.\frac{\partial
					\mathcal{L}}{\partial
					h_i}\right|_{h=0}\,h_i
+\sum_{ij}\left. \frac{1}{2}\frac{\partial^2 \mathcal{L}}{\partial h_i
	   \partial h_j}\right|_{h=0}\,h_i h_j\nonumber\\
& &+\mathcal O(h^3).
\label{likelihood}
\end{eqnarray}
The first and second derivatives are given by
\begin{eqnarray}
\frac{\partial \mathcal{L}}{\partial h_i}&=&{\mathcal H}_i -{\langle {\mathcal H}_i \rangle},\label{L_der}\\
\frac{\partial^2 \mathcal{L}}{\partial h_i \partial h_j}&=&-\frac{1}{2}\mathrm{Tr}\left[\mathbf C^{-1}\frac{\partial \mathbf C}{\partial h_i}\mathbf C^{-1}\frac{\partial \mathbf C}{\partial h_j}\right],
\end{eqnarray}
where
${\mathcal H}_i\equiv \frac{1}{2}\left [ {\bf C^{-1}} {\bm a}
\right]^{\dagger} \frac{\partial \bf C}{\partial h_i} \left [{\bf
C^{-1}} {\bm a} \right]$, and $\bm a$ denotes $a_{lm}$ measured from the
data and ${\mathbf C}\equiv \langle{\bm a}{\bm
a}^\dagger\rangle$, both of which include noise and the other
data-specific terms. 

We obtain an estimator for $\mathbf h$ by maximizing the likelihood with
respect to $\mathbf h$ \cite{Anisotropy_Estimator} 
\begin{eqnarray} 
\hat h_i&=&\sum_j\left[{\mathcal F}^{-1}\right]_{ij} ({\mathcal H}_j-\langle{\mathcal H}_j\rangle),\\
{\mathcal F}_{ij}&\equiv&\frac{1}{2}\mathrm{Tr}\left[\mathbf C^{-1}\frac{\partial \mathbf C}{\partial h_i}\mathbf C^{-1}\frac{\partial \mathbf C}{\partial h_j}\right].
\end{eqnarray}
The covariance matrix, $\mathbf C$, is neither diagonal in pixel nor
harmonic space. In order to reduce the computational cost, we shall 
approximate it as diagonal in harmonic space. While this approximation
makes our estimator sub-optimal, it remains un-biased.
The new estimator is
\begin{eqnarray} 
\hat h_i&=&\frac{1}{2}\sum_{j}\left(\mathbf F^{-1}\right)_{ij}\label{quad_estimator}\\
&\times&\sum_{l_1m_1}\sum_{l_2m_2}\frac{\partial \mathcal C_{l_1m1,l_2m_2}}{\partial h_j} \frac{\tilde a^*_{l_1m_1} \tilde a_{l_2m_2}-\langle \tilde a^*_{l_1m_1} \tilde a_{l_2m_2}\rangle_{h=0}}{(C_{l_1}+N_{l_1})(C_{l_2}+N_{l_2})},\nonumber
\end{eqnarray}
where $\tilde a_{lm}\equiv\int d^2\hat{\mathbf n}~T(\hat{\mathbf
n})M(\hat{\mathbf n})Y_{lm}^*(\hat{\mathbf n})$ is the spherical
harmonic coefficients computed from a masked temperature map
($M(\hat{\mathbf n})=0$ in the masked pixels, and 1 otherwise), and
$C_l$ and $N_l$ are the signal and noise power spectra,
respectively. The matrix ${\mathbf F}$ is defined by
\begin{eqnarray} 
F_{ij}\equiv\frac{f^2_{\mathrm{sky}}}{2}\sum_{l_1m_1}\sum_{l_2m_2}
\frac1{C_{l_1}+N_{l_1}} \frac{\partial \mathcal C_{l_1m1,l_2m_2}}{\partial h_i}
\nonumber\\
\times\frac1{C_{l_2}+N_{l_2}}\frac{\partial \mathcal C_{l_1m1,l_2m_2}}{\partial
 h_j}, 
\end{eqnarray}
with $f_{\mathrm{sky}}\equiv \int \frac{d^2\hat{\bf n}}{4\pi}M(\hat{\bf
n})$ the fraction of unmasked pixels.
Here, $\langle\tilde a^*_{l_1m_1} \tilde a_{l_2m_2}\rangle_{h=0}$ in
Eq.~\ref{quad_estimator} is the ``mean
field,'' which is non-zero even when $g_*=0$.
Data-specific issues such as an incomplete sky coverage,
inhomogeneous noise, and asymmetric beams generate the mean field. 

From $\hat h_i$, we need to estimate $g_*$ and $\hat{\mathbf E}_{\mathrm{cl}}$.
As the estimator $\hat h_i$ consists of the sum of many pairs of
coefficients $a_{lm}$, we expect the estimated value to follow a Gaussian distribution (the central limit theorem). Therefore, the
likelihood of $g_*$ and $\hat{\mathbf E}_{\mathrm{cl}}$ is
\begin{eqnarray} 
\mathcal L&=&\frac{1}{|(2\pi)\mathbf G|^{1/2}}\label{h_like}\\
&\times&\exp\left\{-\frac{1}{2} \left[\hat{\mathbf h}-{\mathbf h}(g_{*},\hat{\mathbf E}_{\mathrm{cl}})\right]^T \mathbf G^{-1} \left[\hat {\mathbf h}-{\mathbf h}(g_{*},\hat{\mathbf E}_{\mathrm{cl}})\right]\right\}\nonumber,
\end{eqnarray}
where $\mathbf G$ is the covariance matrix of $\hat{\mathbf h}$, which we compute from 1000 Monte Carlo simulations. Since ${\mathbf h}(g_{*},\hat{\mathbf E}_{\mathrm{cl}})$ has nonlinear dependence on $g_{*}$ and $\hat{\mathbf E}_{\mathrm{cl}}$, we obtained the posterior distribution of $g_*$ and $\hat{\mathbf E}_{\mathrm{cl}}$ by evaluating Eq.~\ref{h_like} with the Markov Chain Monte Carlo sampling \cite{CosmoMC}.\footnote{One can calculate the expected error bars on
$g_{2M}$ using the Fisher matrix \cite{CMB_anistropic_power}. While such
simplified calculations predict the same error bars on all components of
$g_{2M}$, the actual error bars depend on $M$ due to
the shape of the mask. Also, the Fisher calculations assume homogeneous
noise. Nonetheless, our error bars on $g_{2M}$ from Monte Carlo
simulations and our own Fisher calculations assuming homogeneous noise
and the sky fraction of 71\% are in agreement, to within 20\%.}

\begin{figure*}[t]
\centering\includegraphics[width=0.25\textheight]{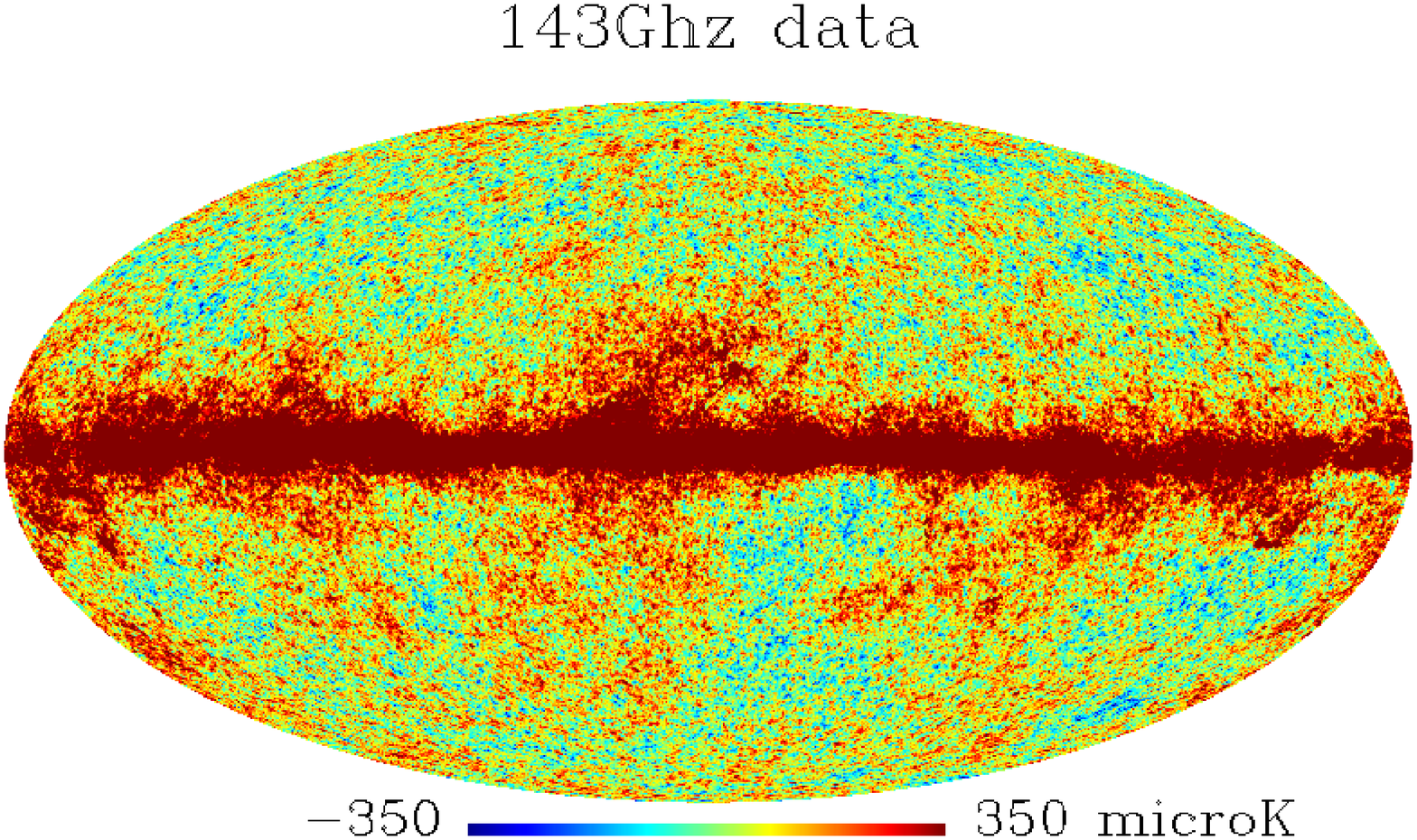}
\centering\includegraphics[width=0.25\textheight]{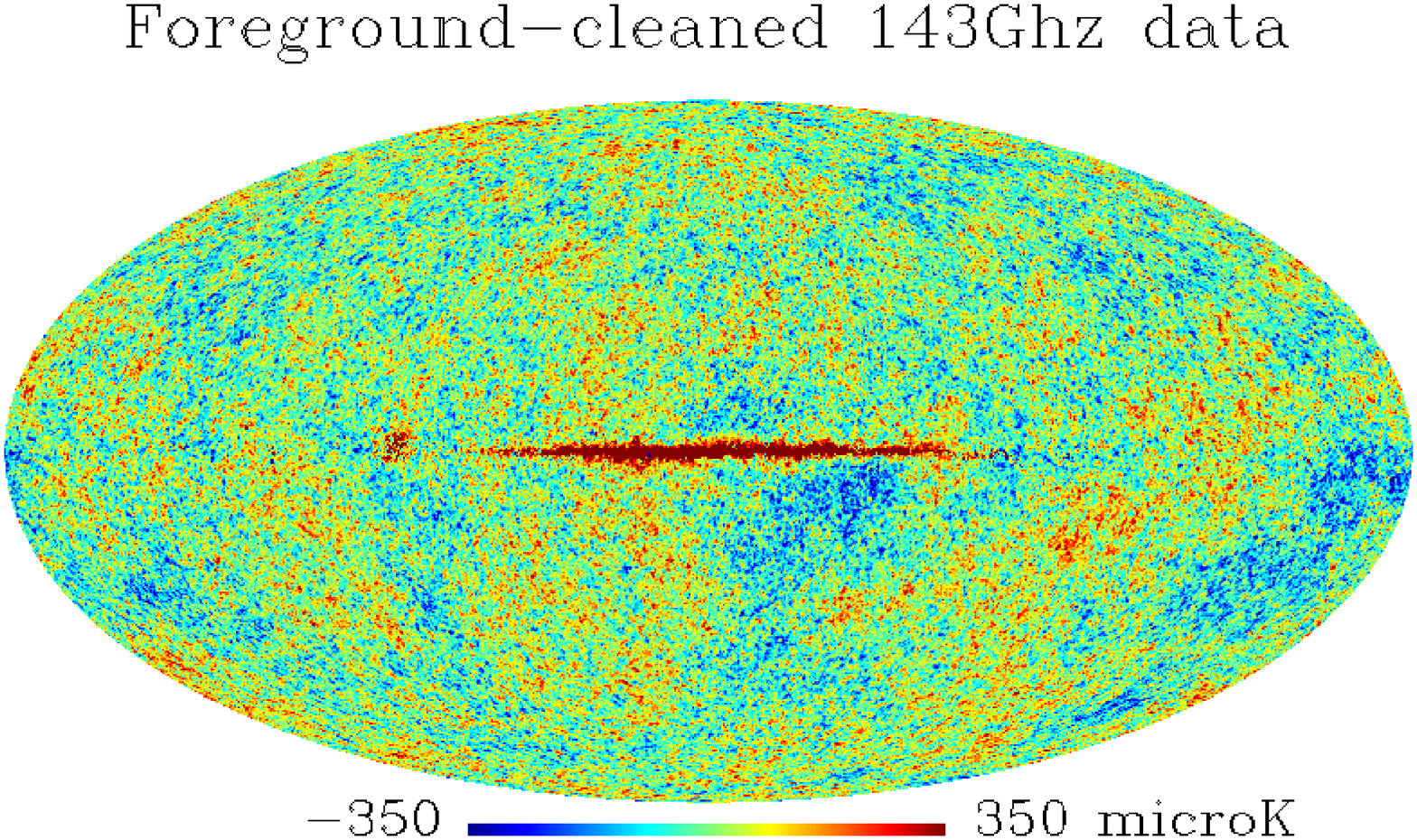}
\centering\includegraphics[width=0.25\textheight]{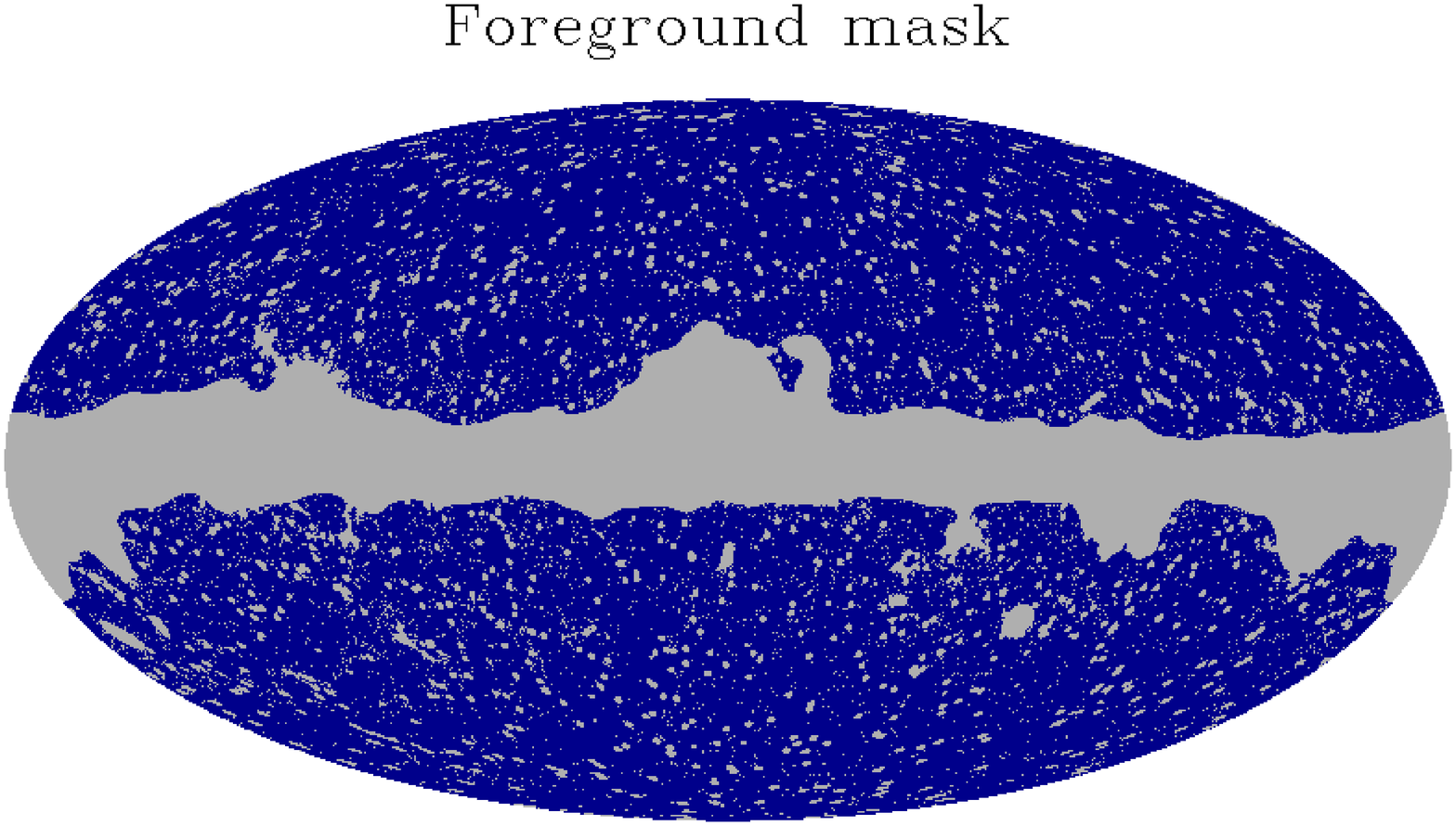}
\caption{(Left) The \textit{Planck} temperature map at 143~GHz. (Middle)
 The foreground-reduced map at  143~GHz. (Right) The foreground
 mask. The maps are shown in a Mollweide projection in Galactic coordinates.}
\label{map}
\end{figure*}
\begin{figure*}[t]
\centering\includegraphics[width=0.25\textheight]{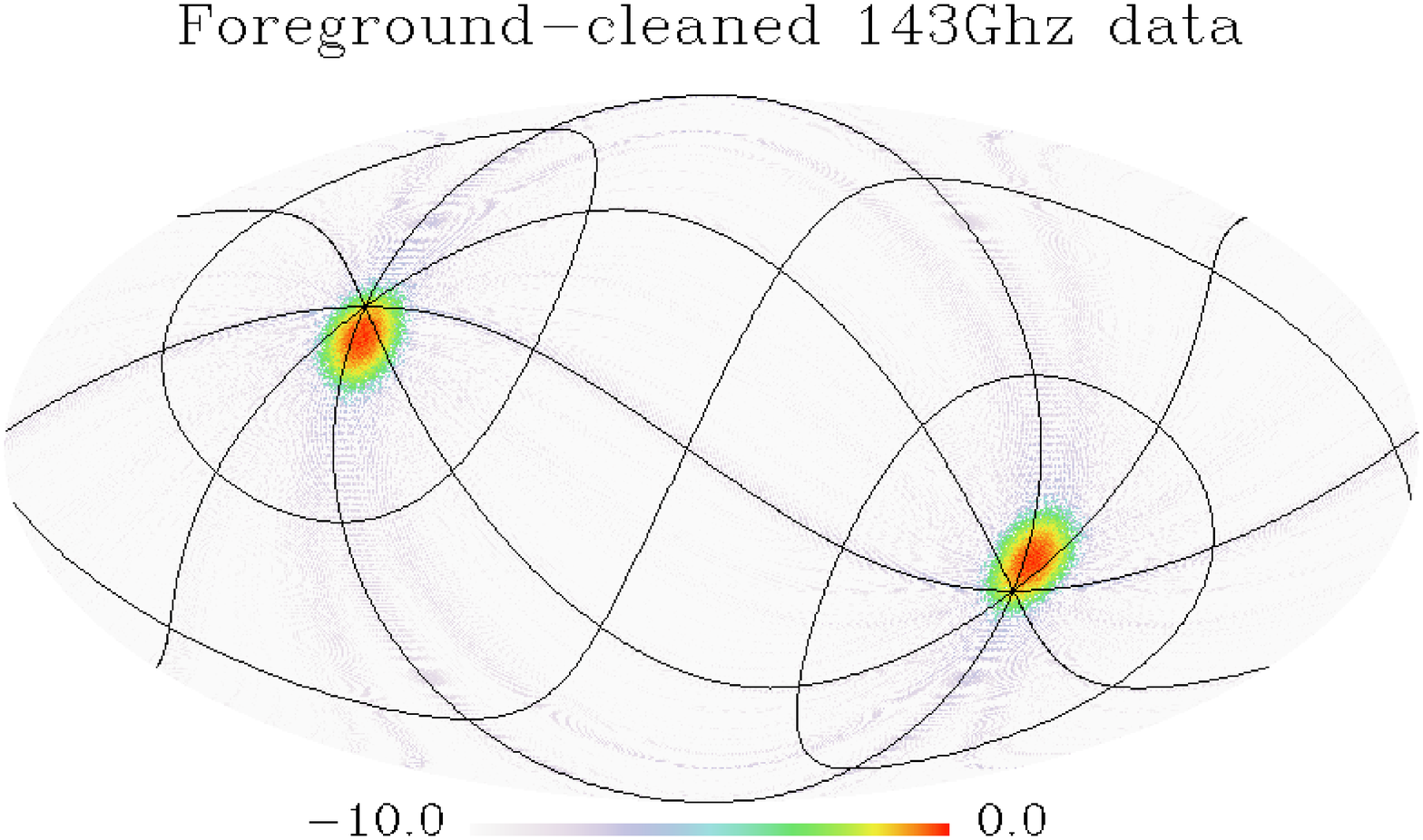}
\centering\includegraphics[width=0.25\textheight]{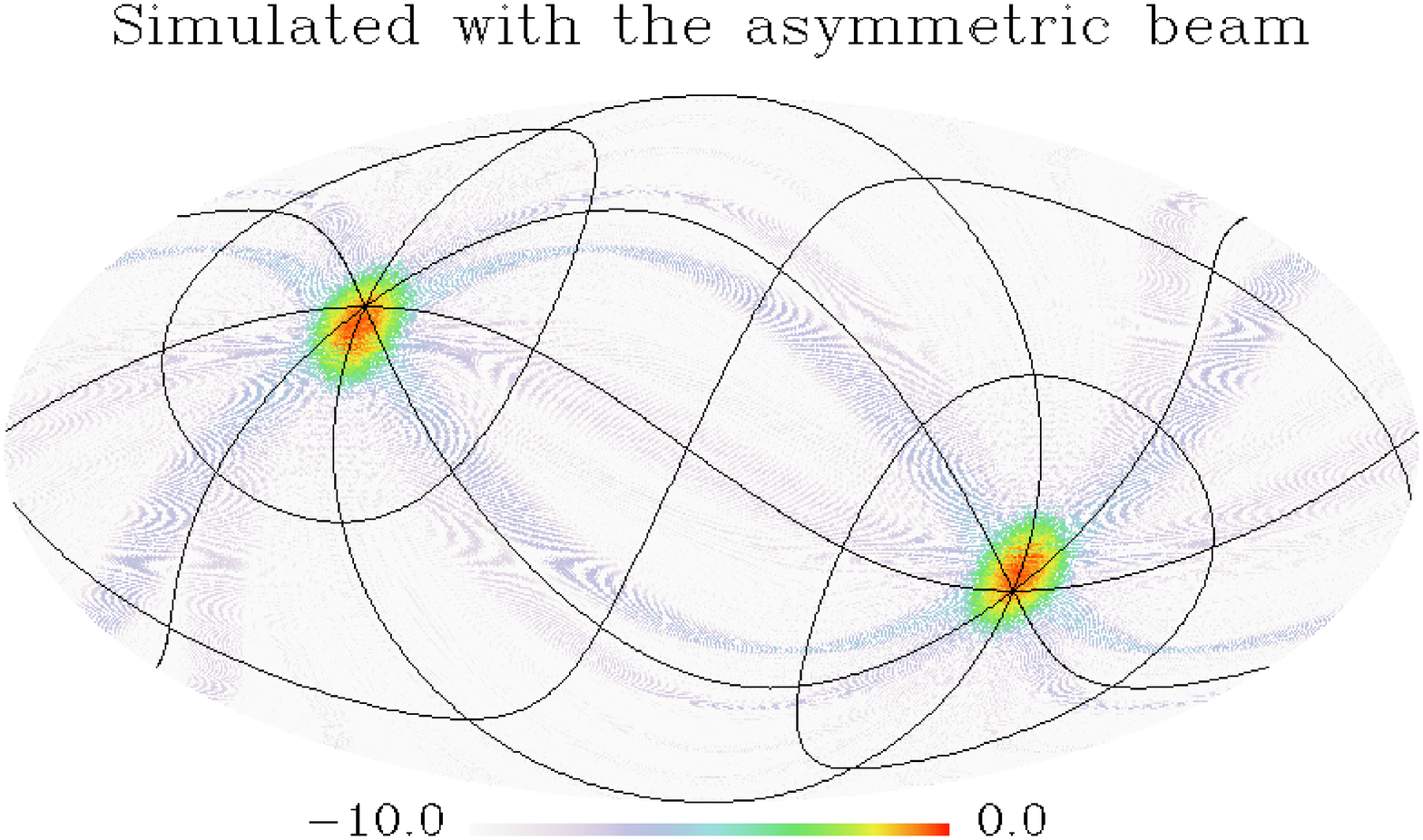}
\centering\includegraphics[width=0.25\textheight]{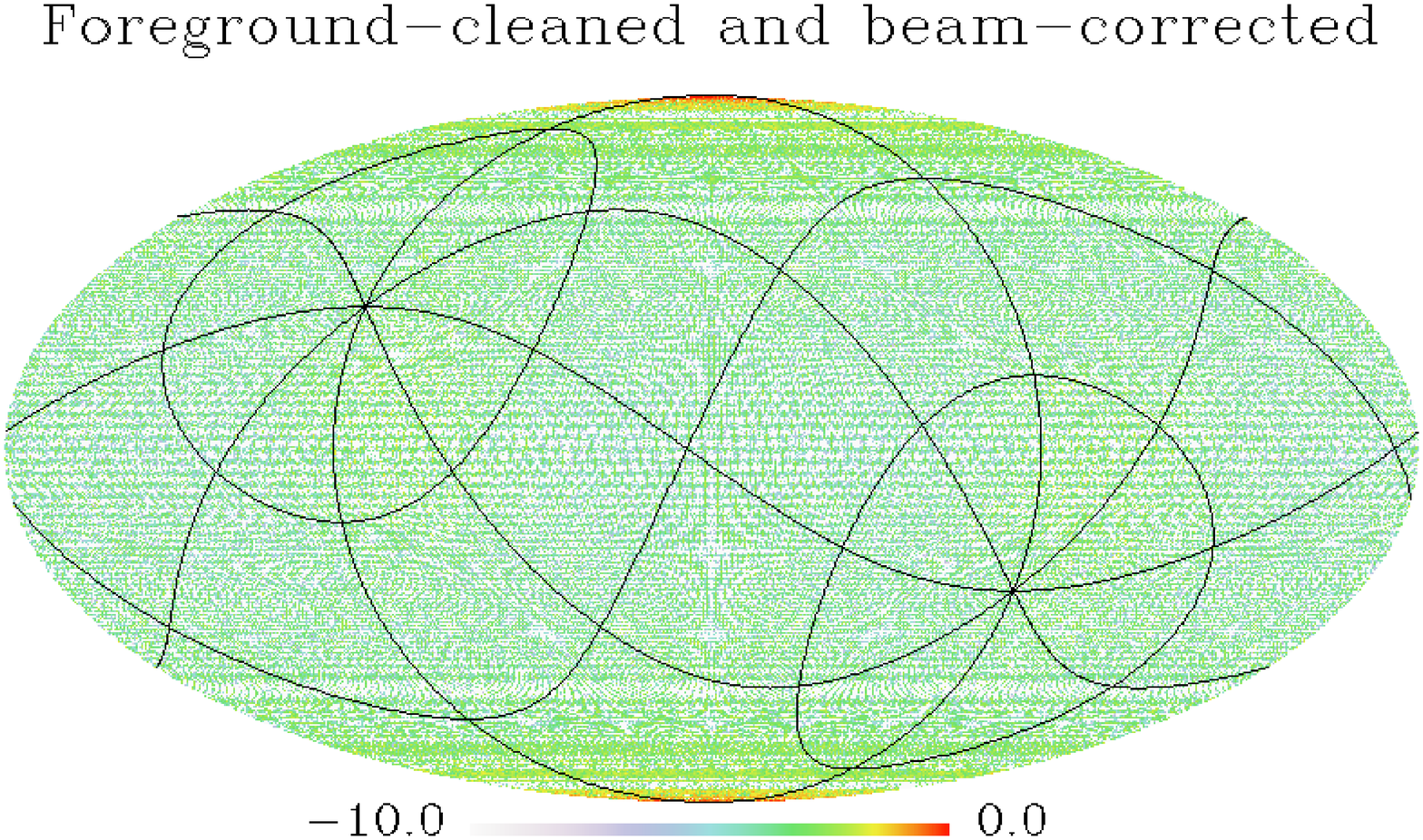}
\caption{(Left) Log-likelihood of locations of a preferred direction,
 $\ln \mathcal L(\hat{\mathbf E}_{\mathrm{cl}})$, 
 computed from the foreground-reduced map at 143~GHz. (Middle) $\ln
 \mathcal L(\hat{\mathbf E}_{\mathrm{cl}})$ from the average of
 simulations with the 
 asymmetric beam. There are two peaks due to parity symmetry. The peaks
 lie close to the Ecliptic pole. The over-laid grids show Ecliptic
 coordinates. (Right) $\ln \mathcal 
 L(\hat{\mathbf E}_{\mathrm{cl}})$ after removing the mean field due to
 the asymmetric  beam. No obvious peaks are left.}
\label{direction}
\end{figure*}

We use the \textit{Planck} 2013 temperature maps at $N_{\rm side}=2048$, which are available at the Planck Legacy Archive
\cite{Planck_overview,Planck_LFI_processing,Planck_HFI_processing}. (We
upgrade the low-frequency maps, which are originally at $N_{\rm
side}=1024$, to $N_{\rm side}=2048$.) We use the map at 143~GHz as the
main ``CMB channel'', and use the other frequencies as ``foreground
templates''. We reduce the diffuse Galactic foreground emission by
fitting templates to, and removing them from, the 143~GHz
map. This is similar to the method called \texttt{SEVEM} by the
\textit{Planck} collaboration \cite{Planck_comp}. We derive the
 templates by taking a difference between two maps at
neighboring frequencies. This procedure ensures the absence of CMB in
the derived templates, producing five templates: $(30-44)$,
$(44-70)$, $(353-217)$, $(545-353)$, and $(857-545)$~[GHz]. To create
these difference maps, we first smooth a pair of maps to the common
resolution. We smooth the low-frequency maps at 30--70~GHz as
$a^{(\nu)}_{lm}\to a^{(\nu)}_{lm}b^{G}_l/b_l^{(\nu)}$, where 
$b_l^{(\nu)}$ is the beam transfer function at a frequency $\nu$
\cite{Planck_LFI_beam} and $b_l^G$ is a Gaussian beam of $33'$
(FWHM). We smooth the high-frequency maps at 217--857~GHz as
$a^{(\nu)}_{lm}\to a^{(\nu)}_{lm}b^{(143)}_l/b_l^{(\nu)}$, where
$b_l^{(143)}$ is the beam transfer function at 143~GHz
\cite{Planck_HFI_beam}.

After the smoothing, we mask the locations of point sources and the
brightest region near the Galactic center (3\% of the sky) following
\texttt{SEVEM} \cite{Planck_comp}. As the smoothed sources occupy
more pixels, we enlarge the original point-source mask as follows: we
create a map having $1$ at the source locations and 0 otherwise, and smooth
it. We then mask the pixels whose values exceed $e^{-2}$. We fit the
templates to the 143~GHz map on the unmasked pixels (86\% of the sky).

The left and middle panels of Figure \ref{map} show the original and
foreground-reduced maps at 143~GHz, respectively. We still find
significant foreground emission on the Galactic plane. We thus mask the
regions contaminated by the residual foreground emission, combining the
masks of various foreground-reduced maps produced by the \textit{Planck}
collaboration (\texttt{NILC}, \texttt{Ruler}, \texttt{SEVEM}, and
\texttt{SMICA} \cite{Planck_comp}), and the point-source mask. We show
the combined mask in the right panel of Figure~\ref{map}, which leaves
71\% of the sky unmasked, and is similar to the ``union mask'' of the
\textit{Planck} collaboration, except for a slightly enlarged
point-source mask due to smoothing.

We use Eqs.~\ref{quad_estimator} and \ref{h_like} to compute $g_{LM}$
from the masked 
foreground-reduced map. We restrict our analysis to the multipole range
of $2\le \ell \le 2000$. We compute the mean field from 1000 Monte-Carlo
realizations of signal and noise. The signal map is $T_S(\hat{\bf
n})=\sum_{lm}\sqrt{C_l}x_{lm}b_l^{(\nu)}p_lY_{lm}(\hat{\bf n})$, where
$C_l$ is the best-fit ``Planck+WP'' power spectrum
\cite{Planck_cosmology}, $p_l$ the pixel window function, and $x_{lm}$ a
Gaussian random variable with unit variance. The noise map is
$T_N(\hat{\bf n})=\sqrt{N(\hat{\bf n})}y(\hat{\bf n})$, where
$N(\hat{\bf n})$ is the noise variance map provided by the
\textit{Planck} collaboration, and $y(\hat{\bf n})$ a Gaussian random
variable with unit variance. We create high-frequency maps at $N_{\rm
side}=2048$, while we create low-frequency maps at $N_{\rm side}=1024$
and upgrade to $N_{\rm side}=2048$. We also compute $g_{LM}$ from the
signal plus noise simulations, and compute the covariance matrix, ${\bf
G}$, in Eq.~\ref{h_like}. Finally, we compute the posterior distribution
of $g_*$ and $\hat{\bf E}_{\rm cl}$ by evaluating Eq.~\ref{h_like} using
the \texttt{CosmoMC} sampler \cite{CosmoMC}.

The left panel of Figure~\ref{direction} shows the log-likelihood of
locations of a preferred direction,
$\ln\mathcal L(\hat{\mathbf E}_{\mathrm{cl}})$, given the \textit{Planck} data. We find
a significant detection of $g_*=-0.111\pm 0.013$ (68\%~CL) with
$\hat{\mathbf E}_{\mathrm{cl}}$ pointing to
$(l,b)=(94^\circ.0^{+3^\circ.9}_{-4^\circ.0},23^\circ.3\pm 4^\circ.1)$ in
Galactic coordinates. This direction lies close to the Ecliptic pole at
$(l,b)=(96^\circ.4,29^\circ.8)$. 

This is essentially the same result as found from the \textit{WMAP}
data. Following the first detection reported in
Ref.~\cite{WMAP5_Eriksen_gstar1}, the subsequent analysis finds
$g_*=0.29\pm 0.031$ with $(l,b)=(94^\circ,26^\circ)\pm 4^\circ$ from the
\textit{WMAP} 5-year map at 94~GHz in the multipole range of
$2\le \ell \le 400$ \cite{WMAP5_Eriksen_gstar2} (also see
\cite{Anisotropy_Estimator}). They find a negative value at 41~GHz,
$g_*=-0.18\pm 0.04$. These signals, however, have been explained
entirely by the effect of \textit{WMAP}'s asymmetric beams coupled with
the scan pattern \cite{gstar_beam,WMAP9:results}. To confirm their
results, we use the foreground-reduced \textit{WMAP} 9-year maps
\cite{WMAP9:results}, finding $g_*=-0.484^{+0.021}_{-0.023}$, $0.105^{+0.036}_{-0.028}$, and $0.355^{+0.038}_{-0.037}$ at 41, 61, and 94~GHz,
respectively, in the multipole range of $2\le \ell \le 1000$. The directions
lie close to the Ecliptic pole.

We find $g_*<0$ from the \textit{Planck} 143 GHz map. This is because
the orientations of the semi-major axes of 143~GHz beams are nearly
parallel to \textit{Planck}'s scan direction \cite{Planck_HFI_beam},
which lies approximately along the Ecliptic longitudes. As the beams are
fatter along the Ecliptic longitudes, the \textit{Planck} measures less
power along the Ecliptic north-south direction than the east-west direction,
yielding a quadrupolar power modulation with $g_*<0$.\footnote{While
\textit{WMAP} does not scan along the Ecliptic longitudes, the
scan directions cover only about 30\% of possible angles on the Ecliptic
equator, which are closer to being parallel to the Ecliptic
longitudes. As a result, the 41~GHz maps give $g_*<0$, as the
orientations of the 41~GHz beams are nearly parallel to \textit{WMAP}'s
scan direction, whereas the 61 and 94~GHz maps give $g_*>0$,
as the orientations are nearly perpendicular to the scan direction
\cite{Hill:2008}. This explanation is due to Ref.~\cite{gstar_beam}.} 

We quantify and remove the effect of beam asymmetry by computing
$g_{LM}$ from 1000 signal plus noise simulations, in which the signal is
convolved with \textit{Planck}'s asymmetric beams and scans. We
have used the \texttt{EffConv} code, which is developed by the \textit{Planck}
collaboration and publicly available\footnote{\url{http://irsa.ipac.caltech.edu/data/Planck/release_1/software}} with the Planck effective beam data files \cite{effConv,Planck_HFI_beam}. The middle panel of
Figure~\ref{direction} shows $\ln\mathcal L(\hat{\mathbf E}_{\mathrm{cl}})$ given the
simulation data. We reproduce what we find from the real data:
$g_*=-0.101\pm 0.0004$ with $(96^\circ.1\pm 0^\circ.1, 25^\circ.9\pm 0^\circ.1)$
(the error bars are for the average of simulations). Using this
result as the mean field (i.e., $\langle \tilde a^*_{l_1m_1} \tilde a_{l_2m_2}\rangle_{h=0}$ in Eq. \ref{quad_estimator}), we  recompute $\ln\mathcal
L(g_*,\hat{\mathbf E}_{\mathrm{cl}})$, finding no evidence for  $g_*$ (see also the right panel of Figure~\ref{direction}, which shows no preferred direction). Our best limit is $g_*=0.002\pm 0.016$ (68\%~CL), $0.002^{+0.031}_{-0.032}$ (95\%~CL) and $0.002^{+0.047}_{-0.048}$ (99.7\%~CL).

We have also analyzed the foreground-reduced 100~GHz map, which has less
foreground emission than the 143~GHz map. We find 28- and 7-$\sigma$
detections of $g_*$ in the Ecliptic-pole directions before and after the
beam asymmetry correction, respectively. The 100~GHz beam is much less
symmetric than the 143~GHz one \cite{Planck_HFI_beam}; thus, the beam
simulation needs to be more precise for removing the asymmetry to the
sufficient level. We find $g_*=-0.308\pm 0.011$ before the beam
asymmetry correction, which is consistent with the 100~GHz beams being
more elongated along \textit{Planck}'s scan direction.

Finally, we study the effect of Galactic foreground emission. Using the
{\it raw} 143~GHz without cleaning, we find significant anisotropy:
$g_*=0.340$ and $0.328\pm 0.018$ before and after the beam asymmetry
correction, respectively. The directions lie close to the Galactic pole;
thus, the foreground reduction plays an important role in nulling
artificial anisotropy in the data.

\begin{table}[h!]
\centering
\caption{Best-fit amplitudes and directions with the 68\%~CL
 intervals. ``BC'' and ``FR'' stand for ``Beam Correction''   
 and ``Foreground Reduction,'' respectively. The last row shows the
 result from the average of 1000 asymmetric beam simulations.} 
\begin{tabular}{ccccc}
\hline\hline 
BC & FR &$g_*$ & direction $(l,b)$ [degrees] \\
\hline
No & No &$\;\;\;0.340\pm 0.018$ & $(226.6^{+21.2}_{-24.3},  85.8\pm 1.5)$\\ 
Yes & No &$\;\;\;0.328\pm 0.018$ & $(141.1^{+18.6}_{-19.7}, 85.3\pm 1.8)$\\ 
No & Yes & $ -0.111\pm 0.013$ & $\;\;\;\;(94.0^{+3.9}_{-4.0},\;23.3\pm 4.1)$\\ 
\textbf{Yes} & \textbf{Yes}& $\;\;\; \mathbf{0.002\pm 0.016}$ & $\mathbf{(180.7^{+179.3}_{-180.7},44.8^{+45.2}_{-44.8})}$\\ 
\hline
No & ---  & $\;-0.101\pm 0.0004$ & $(96.1\pm 0.1, 25.9\pm 0.1)$\\ 
\hline
\end{tabular}
\label{gstar}
\end{table}
We summarize our finding in Table \ref{gstar}. After removing the 
effects of \textit{Planck}'s asymmetric beams and Galactic
foreground emission, we find no evidence for $g_*$. Our limit, about 2\%
in $g_*$, provides the most stringent test of rotational symmetry during
inflation.\\\\

JK would like to thank Belen Barreiro, Carlo Baccigalupi, Jacques
Delabrouille, Sanjit Mitra, Anthony Lewis and Niels Oppermann for helpful discussions. 
We acknowledge the use of the Planck Legacy Archive (PLA). The development of \textit{Planck} has been supported by: ESA; CNES and CNRS/INSU-IN2P3-INP (France); ASI, CNR, and INAF (Italy); NASA and DoE (USA); STFC and UKSA (UK); CSIC, MICINN and JA (Spain); Tekes, AoF and CSC (Finland); DLR and MPG (Germany); CSA (Canada); DTU Space (Denmark); SER/SSO (Switzerland); RCN (Norway); SFI (Ireland); FCT/MCTES (Portugal); and PRACE (EU). 
A description of the Planck Collaboration and a list of its members, including the technical or scientific activities in which they have been involved, can be found at \url{http://www.sciops.esa.int/index.php?project=planck&page=Planck_Collaboration}.
We acknowledge the use of the Legacy Archive for Microwave Background
Data Analysis (LAMBDA), part of the High Energy Astrophysics Science
Archive Center (HEASARC). HEASARC/LAMBDA is a service of the
Astrophysics Science Division at the NASA Goddard Space Flight Center.
We also acknowledge the use of the \texttt{EffConv} \cite{effConv},
\texttt{HEALPix} \cite{HEALPix:framework}, \texttt{CAMB} \cite{CAMB}, and
\texttt{CosmoMC} packages \cite{CosmoMC}.  
\bibliographystyle{apsrev4-1}
\bibliography{bibliography}

\begin{thebibliography}{36}%
\makeatletter
\providecommand \@ifxundefined [1]{%
 \@ifx{#1\undefined}
}%
\providecommand \@ifnum [1]{%
 \ifnum #1\expandafter \@firstoftwo
 \else \expandafter \@secondoftwo
 \fi
}%
\providecommand \@ifx [1]{%
 \ifx #1\expandafter \@firstoftwo
 \else \expandafter \@secondoftwo
 \fi
}%
\providecommand \natexlab [1]{#1}%
\providecommand \enquote  [1]{``#1''}%
\providecommand \bibnamefont  [1]{#1}%
\providecommand \bibfnamefont [1]{#1}%
\providecommand \citenamefont [1]{#1}%
\providecommand \href@noop [0]{\@secondoftwo}%
\providecommand \href [0]{\begingroup \@sanitize@url \@href}%
\providecommand \@href[1]{\@@startlink{#1}\@@href}%
\providecommand \@@href[1]{\endgroup#1\@@endlink}%
\providecommand \@sanitize@url [0]{\catcode `\\12\catcode `\$12\catcode
  `\&12\catcode `\#12\catcode `\^12\catcode `\_12\catcode `\%12\relax}%
\providecommand \@@startlink[1]{}%
\providecommand \@@endlink[0]{}%
\providecommand \url  [0]{\begingroup\@sanitize@url \@url }%
\providecommand \@url [1]{\endgroup\@href {#1}{\urlprefix }}%
\providecommand \urlprefix  [0]{URL }%
\providecommand \Eprint [0]{\href }%
\providecommand \doibase [0]{http://dx.doi.org/}%
\providecommand \selectlanguage [0]{\@gobble}%
\providecommand \bibinfo  [0]{\@secondoftwo}%
\providecommand \bibfield  [0]{\@secondoftwo}%
\providecommand \translation [1]{[#1]}%
\providecommand \BibitemOpen [0]{}%
\providecommand \bibitemStop [0]{}%
\providecommand \bibitemNoStop [0]{.\EOS\space}%
\providecommand \EOS [0]{\spacefactor3000\relax}%
\providecommand \BibitemShut  [1]{\csname bibitem#1\endcsname}%
\let\auto@bib@innerbib\@empty
\bibitem [{\citenamefont {Starobinsky}(1980)}]{Starobinsky:1980}%
  \BibitemOpen
  \bibfield  {author} {\bibinfo {author} {\bibfnamefont {A.~A.}\ \bibnamefont
  {Starobinsky}},\ }\href {\doibase 10.1016/0370-2693(80)90670-X} {\bibfield
  {journal} {\bibinfo  {journal} {Phys.Lett.}\ }\textbf {\bibinfo {volume}
  {B91}},\ \bibinfo {pages} {99} (\bibinfo {year} {1980})}\BibitemShut
  {NoStop}%
\bibitem [{\citenamefont {Sato}(1981)}]{Sato:1980}%
  \BibitemOpen
  \bibfield  {author} {\bibinfo {author} {\bibfnamefont {K.}~\bibnamefont
  {Sato}},\ }\href@noop {} {\bibfield  {journal} {\bibinfo  {journal}
  {Mon.Not.Roy.Astron.Soc.}\ }\textbf {\bibinfo {volume} {195}},\ \bibinfo
  {pages} {467} (\bibinfo {year} {1981})}\BibitemShut {NoStop}%
\bibitem [{\citenamefont {Guth}(1981)}]{Guth:1980}%
  \BibitemOpen
  \bibfield  {author} {\bibinfo {author} {\bibfnamefont {A.~H.}\ \bibnamefont
  {Guth}},\ }\href {\doibase 10.1103/PhysRevD.23.347} {\bibfield  {journal}
  {\bibinfo  {journal} {Phys.Rev.}\ }\textbf {\bibinfo {volume} {D23}},\
  \bibinfo {pages} {347} (\bibinfo {year} {1981})}\BibitemShut {NoStop}%
\bibitem [{\citenamefont {Linde}(1982)}]{Linde:1981}%
  \BibitemOpen
  \bibfield  {author} {\bibinfo {author} {\bibfnamefont {A.~D.}\ \bibnamefont
  {Linde}},\ }\href {\doibase 10.1016/0370-2693(82)91219-9} {\bibfield
  {journal} {\bibinfo  {journal} {Phys.Lett.}\ }\textbf {\bibinfo {volume}
  {B108}},\ \bibinfo {pages} {389} (\bibinfo {year} {1982})}\BibitemShut
  {NoStop}%
\bibitem [{\citenamefont {Albrecht}\ and\ \citenamefont
  {Steinhardt}(1982)}]{Albrecht:1982}%
  \BibitemOpen
  \bibfield  {author} {\bibinfo {author} {\bibfnamefont {A.}~\bibnamefont
  {Albrecht}}\ and\ \bibinfo {author} {\bibfnamefont {P.~J.}\ \bibnamefont
  {Steinhardt}},\ }\href {\doibase 10.1103/PhysRevLett.48.1220} {\bibfield
  {journal} {\bibinfo  {journal} {Phys.Rev.Lett.}\ }\textbf {\bibinfo {volume}
  {48}},\ \bibinfo {pages} {1220} (\bibinfo {year} {1982})}\BibitemShut
  {NoStop}%
\bibitem [{\citenamefont {Mukhanov}\ and\ \citenamefont
  {Chibisov}(1981)}]{Mukhanov:1981}%
  \BibitemOpen
  \bibfield  {author} {\bibinfo {author} {\bibfnamefont {V.~F.}\ \bibnamefont
  {Mukhanov}}\ and\ \bibinfo {author} {\bibfnamefont {G.}~\bibnamefont
  {Chibisov}},\ }\href@noop {} {\bibfield  {journal} {\bibinfo  {journal} {JETP
  Lett.}\ }\textbf {\bibinfo {volume} {33}},\ \bibinfo {pages} {532} (\bibinfo
  {year} {1981})}\BibitemShut {NoStop}%
\bibitem [{\citenamefont {{Hinshaw}}\ \emph {et~al.}(2013)\citenamefont
  {{Hinshaw}} \emph {et~al.}}]{WMAP9:cosmology}%
  \BibitemOpen
  \bibfield  {author} {\bibinfo {author} {\bibfnamefont {G.}~\bibnamefont
  {{Hinshaw}}} \emph {et~al.},\ }\href {\doibase 10.1088/0067-0049/208/2/19}
  {\bibfield  {journal} {\bibinfo  {journal} {\apjs}\ }\textbf {\bibinfo
  {volume} {208}},\ \bibinfo {eid} {19} (\bibinfo {year} {2013})},\ \bibinfo
  {note} {arXiv:1212.5226}\BibitemShut {NoStop}%
\bibitem [{\citenamefont {Ade}\ \emph {et~al.}(2013{\natexlab{a}})\citenamefont
  {Ade} \emph {et~al.}}]{Planck_cosmology}%
  \BibitemOpen
  \bibfield  {author} {\bibinfo {author} {\bibfnamefont {P.}~\bibnamefont
  {Ade}} \emph {et~al.} (\bibinfo {collaboration} {Planck Collaboration}),\
  }\href@noop {} {\bibfield  {journal} {\bibinfo  {journal} {ArXiv e-prints}\ }
  (\bibinfo {year} {2013}{\natexlab{a}})},\ \Eprint
  {http://arxiv.org/abs/1303.5076} {arXiv:1303.5076 [astro-ph.CO]} \BibitemShut
  {NoStop}%
\bibitem [{\citenamefont {Ackerman}\ \emph {et~al.}(2007)\citenamefont
  {Ackerman}, \citenamefont {Carroll},\ and\ \citenamefont
  {Wise}}]{Ackerman:2007}%
  \BibitemOpen
  \bibfield  {author} {\bibinfo {author} {\bibfnamefont {L.}~\bibnamefont
  {Ackerman}}, \bibinfo {author} {\bibfnamefont {S.~M.}\ \bibnamefont
  {Carroll}}, \ and\ \bibinfo {author} {\bibfnamefont {M.~B.}\ \bibnamefont
  {Wise}},\ }\href {\doibase 10.1103/PhysRevD.75.083502,
  10.1103/PhysRevD.80.069901} {\bibfield  {journal} {\bibinfo  {journal}
  {Phys.Rev.}\ }\textbf {\bibinfo {volume} {D75}},\ \bibinfo {pages} {083502}
  (\bibinfo {year} {2007})},\ \Eprint {http://arxiv.org/abs/astro-ph/0701357}
  {arXiv:astro-ph/0701357 [astro-ph]} \BibitemShut {NoStop}%
\bibitem [{\citenamefont {Soda}(2012)}]{Soda:2012}%
  \BibitemOpen
  \bibfield  {author} {\bibinfo {author} {\bibfnamefont {J.}~\bibnamefont
  {Soda}},\ }\href {\doibase 10.1088/0264-9381/29/8/083001} {\bibfield
  {journal} {\bibinfo  {journal} {Class.Quant.Grav.}\ }\textbf {\bibinfo
  {volume} {29}},\ \bibinfo {pages} {083001} (\bibinfo {year} {2012})},\
  \Eprint {http://arxiv.org/abs/1201.6434} {arXiv:1201.6434 [hep-th]}
  \BibitemShut {NoStop}%
\bibitem [{\citenamefont {Maleknejad}\ \emph {et~al.}(2013)\citenamefont
  {Maleknejad}, \citenamefont {Sheikh-Jabbari},\ and\ \citenamefont
  {Soda}}]{Maleknejad:2012}%
  \BibitemOpen
  \bibfield  {author} {\bibinfo {author} {\bibfnamefont {A.}~\bibnamefont
  {Maleknejad}}, \bibinfo {author} {\bibfnamefont {M.}~\bibnamefont
  {Sheikh-Jabbari}}, \ and\ \bibinfo {author} {\bibfnamefont {J.}~\bibnamefont
  {Soda}},\ }\href {\doibase 10.1016/j.physrep.2013.03.003} {\bibfield
  {journal} {\bibinfo  {journal} {Phys.Rept.}\ }\textbf {\bibinfo {volume}
  {528}},\ \bibinfo {pages} {161} (\bibinfo {year} {2013})},\ \Eprint
  {http://arxiv.org/abs/1212.2921} {arXiv:1212.2921 [hep-th]} \BibitemShut
  {NoStop}%
\bibitem [{\citenamefont {Dimastrogiovanni}\ \emph {et~al.}(2010)\citenamefont
  {Dimastrogiovanni}, \citenamefont {Bartolo}, \citenamefont {Matarrese},\ and\
  \citenamefont {Riotto}}]{Dimastrogiovanni:2010}%
  \BibitemOpen
  \bibfield  {author} {\bibinfo {author} {\bibfnamefont {E.}~\bibnamefont
  {Dimastrogiovanni}}, \bibinfo {author} {\bibfnamefont {N.}~\bibnamefont
  {Bartolo}}, \bibinfo {author} {\bibfnamefont {S.}~\bibnamefont {Matarrese}},
  \ and\ \bibinfo {author} {\bibfnamefont {A.}~\bibnamefont {Riotto}},\ }\href
  {\doibase 10.1155/2010/752670} {\bibfield  {journal} {\bibinfo  {journal}
  {Adv.Astron.}\ }\textbf {\bibinfo {volume} {2010}},\ \bibinfo {pages}
  {752670} (\bibinfo {year} {2010})},\ \Eprint {http://arxiv.org/abs/1001.4049}
  {arXiv:1001.4049 [astro-ph.CO]} \BibitemShut {NoStop}%
\bibitem [{\citenamefont {Barnaby}\ \emph {et~al.}(2012)\citenamefont
  {Barnaby}, \citenamefont {Namba},\ and\ \citenamefont
  {Peloso}}]{Barnaby:2012}%
  \BibitemOpen
  \bibfield  {author} {\bibinfo {author} {\bibfnamefont {N.}~\bibnamefont
  {Barnaby}}, \bibinfo {author} {\bibfnamefont {R.}~\bibnamefont {Namba}}, \
  and\ \bibinfo {author} {\bibfnamefont {M.}~\bibnamefont {Peloso}},\ }\href
  {\doibase 10.1103/PhysRevD.85.123523} {\bibfield  {journal} {\bibinfo
  {journal} {Phys.Rev.}\ }\textbf {\bibinfo {volume} {D85}},\ \bibinfo {pages}
  {123523} (\bibinfo {year} {2012})},\ \Eprint {http://arxiv.org/abs/1202.1469}
  {arXiv:1202.1469 [astro-ph.CO]} \BibitemShut {NoStop}%
\bibitem [{\citenamefont {Shiraishi}\ \emph {et~al.}(2013)\citenamefont
  {Shiraishi}, \citenamefont {Komatsu}, \citenamefont {Peloso},\ and\
  \citenamefont {Barnaby}}]{Shiraishi:2013}%
  \BibitemOpen
  \bibfield  {author} {\bibinfo {author} {\bibfnamefont {M.}~\bibnamefont
  {Shiraishi}}, \bibinfo {author} {\bibfnamefont {E.}~\bibnamefont {Komatsu}},
  \bibinfo {author} {\bibfnamefont {M.}~\bibnamefont {Peloso}}, \ and\ \bibinfo
  {author} {\bibfnamefont {N.}~\bibnamefont {Barnaby}},\ }\href {\doibase
  10.1088/1475-7516/2013/05/002} {\bibfield  {journal} {\bibinfo  {journal}
  {JCAP}\ }\textbf {\bibinfo {volume} {1305}},\ \bibinfo {pages} {002}
  (\bibinfo {year} {2013})},\ \Eprint {http://arxiv.org/abs/1302.3056}
  {arXiv:1302.3056 [astro-ph.CO]} \BibitemShut {NoStop}%
\bibitem [{\citenamefont {{Bartolo}}\ \emph {et~al.}(2013)\citenamefont
  {{Bartolo}}, \citenamefont {{Matarrese}}, \citenamefont {{Peloso}},\ and\
  \citenamefont {{Ricciardone}}}]{Bartolo_F2:2013}%
  \BibitemOpen
  \bibfield  {author} {\bibinfo {author} {\bibfnamefont {N.}~\bibnamefont
  {{Bartolo}}}, \bibinfo {author} {\bibfnamefont {S.}~\bibnamefont
  {{Matarrese}}}, \bibinfo {author} {\bibfnamefont {M.}~\bibnamefont
  {{Peloso}}}, \ and\ \bibinfo {author} {\bibfnamefont {A.}~\bibnamefont
  {{Ricciardone}}},\ }\href {\doibase 10.1103/PhysRevD.87.023504} {\bibfield
  {journal} {\bibinfo  {journal} {\prd}\ }\textbf {\bibinfo {volume} {87}},\
  \bibinfo {eid} {023504} (\bibinfo {year} {2013})},\ \Eprint
  {http://arxiv.org/abs/1210.3257} {arXiv:1210.3257 [astro-ph.CO]} \BibitemShut
  {NoStop}%
\bibitem [{\citenamefont {Ade}\ \emph {et~al.}(2013{\natexlab{b}})\citenamefont
  {Ade} \emph {et~al.}}]{planck2013fnl}%
  \BibitemOpen
  \bibfield  {author} {\bibinfo {author} {\bibfnamefont {P.}~\bibnamefont
  {Ade}} \emph {et~al.} (\bibinfo {collaboration} {Planck Collaboration}),\
  }\href@noop {} {\  (\bibinfo {year} {2013}{\natexlab{b}})},\ \Eprint
  {http://arxiv.org/abs/1303.5084} {arXiv:1303.5084 [astro-ph.CO]} \BibitemShut
  {NoStop}%
\bibitem [{\citenamefont {Schmidt}\ and\ \citenamefont
  {Hui}(2013)}]{Schmidt:2012}%
  \BibitemOpen
  \bibfield  {author} {\bibinfo {author} {\bibfnamefont {F.}~\bibnamefont
  {Schmidt}}\ and\ \bibinfo {author} {\bibfnamefont {L.}~\bibnamefont {Hui}},\
  }\href {\doibase 10.1103/PhysRevLett.110.059902,
  10.1103/PhysRevLett.110.011301} {\bibfield  {journal} {\bibinfo  {journal}
  {Phys.Rev.Lett.}\ }\textbf {\bibinfo {volume} {110}},\ \bibinfo {pages}
  {011301} (\bibinfo {year} {2013})},\ \Eprint {http://arxiv.org/abs/1210.2965}
  {arXiv:1210.2965 [astro-ph.CO]} \BibitemShut {NoStop}%
\bibitem [{\citenamefont {Bartolo}\ \emph {et~al.}(2013)\citenamefont
  {Bartolo}, \citenamefont {Matarrese}, \citenamefont {Peloso},\ and\
  \citenamefont {Ricciardone}}]{Bartolo:2013}%
  \BibitemOpen
  \bibfield  {author} {\bibinfo {author} {\bibfnamefont {N.}~\bibnamefont
  {Bartolo}}, \bibinfo {author} {\bibfnamefont {S.}~\bibnamefont {Matarrese}},
  \bibinfo {author} {\bibfnamefont {M.}~\bibnamefont {Peloso}}, \ and\ \bibinfo
  {author} {\bibfnamefont {A.}~\bibnamefont {Ricciardone}},\ }\href {\doibase
  10.1088/1475-7516/2013/08/022} {\bibfield  {journal} {\bibinfo  {journal}
  {JCAP}\ }\textbf {\bibinfo {volume} {1308}},\ \bibinfo {pages} {022}
  (\bibinfo {year} {2013})},\ \Eprint {http://arxiv.org/abs/1306.4160}
  {arXiv:1306.4160 [astro-ph.CO]} \BibitemShut {NoStop}%
\bibitem [{\citenamefont {{Ma}}\ \emph {et~al.}(2011)\citenamefont {{Ma}},
  \citenamefont {{Efstathiou}},\ and\ \citenamefont
  {{Challinor}}}]{gstar_forecast}%
  \BibitemOpen
  \bibfield  {author} {\bibinfo {author} {\bibfnamefont {Y.-Z.}\ \bibnamefont
  {{Ma}}}, \bibinfo {author} {\bibfnamefont {G.}~\bibnamefont {{Efstathiou}}},
  \ and\ \bibinfo {author} {\bibfnamefont {A.}~\bibnamefont {{Challinor}}},\
  }\href {\doibase 10.1103/PhysRevD.83.083005} {\bibfield  {journal} {\bibinfo
  {journal} {\prd}\ }\textbf {\bibinfo {volume} {83}},\ \bibinfo {pages}
  {083005} (\bibinfo {year} {2011})},\ \Eprint
  {http://arxiv.org/abs/arXiv:1102.4961} {arXiv:1102.4961} \BibitemShut
  {NoStop}%
\bibitem [{\citenamefont {{Hanson}}\ and\ \citenamefont
  {{Lewis}}(2009)}]{Anisotropy_Estimator}%
  \BibitemOpen
  \bibfield  {author} {\bibinfo {author} {\bibfnamefont {D.}~\bibnamefont
  {{Hanson}}}\ and\ \bibinfo {author} {\bibfnamefont {A.}~\bibnamefont
  {{Lewis}}},\ }\href {\doibase 10.1103/PhysRevD.80.063004} {\bibfield
  {journal} {\bibinfo  {journal} {\prd}\ }\textbf {\bibinfo {volume} {80}},\
  \bibinfo {pages} {063004} (\bibinfo {year} {2009})},\ \Eprint
  {http://arxiv.org/abs/0908.0963} {arXiv:0908.0963} \BibitemShut {NoStop}%
\bibitem [{\citenamefont {Lewis}\ and\ \citenamefont {Bridle}(2002)}]{CosmoMC}%
  \BibitemOpen
  \bibfield  {author} {\bibinfo {author} {\bibfnamefont {A.}~\bibnamefont
  {Lewis}}\ and\ \bibinfo {author} {\bibfnamefont {S.}~\bibnamefont {Bridle}},\
  }\href@noop {} {\bibfield  {journal} {\bibinfo  {journal} {\prd}\ }\textbf
  {\bibinfo {volume} {66}},\ \bibinfo {pages} {103511} (\bibinfo {year}
  {2002})}\BibitemShut {NoStop}%
\bibitem [{\citenamefont {{Pullen}}\ and\ \citenamefont
  {{Kamionkowski}}(2007)}]{CMB_anistropic_power}%
  \BibitemOpen
  \bibfield  {author} {\bibinfo {author} {\bibfnamefont {A.~R.}\ \bibnamefont
  {{Pullen}}}\ and\ \bibinfo {author} {\bibfnamefont {M.}~\bibnamefont
  {{Kamionkowski}}},\ }\href {\doibase 10.1103/PhysRevD.76.103529} {\bibfield
  {journal} {\bibinfo  {journal} {\prd}\ }\textbf {\bibinfo {volume} {76}},\
  \bibinfo {pages} {103529} (\bibinfo {year} {2007})},\ \Eprint
  {http://arxiv.org/abs/0709.1144} {arXiv:0709.1144} \BibitemShut {NoStop}%
\bibitem [{\citenamefont {Ade}\ \emph {et~al.}(2013{\natexlab{c}})\citenamefont
  {Ade} \emph {et~al.}}]{Planck_overview}%
  \BibitemOpen
  \bibfield  {author} {\bibinfo {author} {\bibfnamefont {P.}~\bibnamefont
  {Ade}} \emph {et~al.} (\bibinfo {collaboration} {Planck Collaboration}),\
  }\href@noop {} {\bibfield  {journal} {\bibinfo  {journal} {ArXiv e-prints}\ }
  (\bibinfo {year} {2013}{\natexlab{c}})},\ \Eprint
  {http://arxiv.org/abs/1303.5062} {arXiv:1303.5062 [astro-ph.CO]} \BibitemShut
  {NoStop}%
\bibitem [{\citenamefont {Aghanim}\ \emph
  {et~al.}(2013{\natexlab{a}})\citenamefont {Aghanim} \emph
  {et~al.}}]{Planck_LFI_processing}%
  \BibitemOpen
  \bibfield  {author} {\bibinfo {author} {\bibfnamefont {N.}~\bibnamefont
  {Aghanim}} \emph {et~al.} (\bibinfo {collaboration} {Planck Collaboration}),\
  }\href@noop {} {\bibfield  {journal} {\bibinfo  {journal} {ArXiv e-prints}\ }
  (\bibinfo {year} {2013}{\natexlab{a}})},\ \Eprint
  {http://arxiv.org/abs/1303.5063} {arXiv:1303.5063 [astro-ph.IM]} \BibitemShut
  {NoStop}%
\bibitem [{\citenamefont {Ade}\ \emph {et~al.}(2013{\natexlab{d}})\citenamefont
  {Ade} \emph {et~al.}}]{Planck_HFI_processing}%
  \BibitemOpen
  \bibfield  {author} {\bibinfo {author} {\bibfnamefont {P.}~\bibnamefont
  {Ade}} \emph {et~al.} (\bibinfo {collaboration} {Planck Collaboration}),\
  }\href@noop {} {\bibfield  {journal} {\bibinfo  {journal} {ArXiv e-prints}\ }
  (\bibinfo {year} {2013}{\natexlab{d}})},\ \Eprint
  {http://arxiv.org/abs/1303.5067} {arXiv:1303.5067 [astro-ph.CO]} \BibitemShut
  {NoStop}%
\bibitem [{\citenamefont {Ade}\ \emph {et~al.}(2013{\natexlab{e}})\citenamefont
  {Ade} \emph {et~al.}}]{Planck_comp}%
  \BibitemOpen
  \bibfield  {author} {\bibinfo {author} {\bibfnamefont {P.}~\bibnamefont
  {Ade}} \emph {et~al.} (\bibinfo {collaboration} {Planck Collaboration}),\
  }\href@noop {} {\bibfield  {journal} {\bibinfo  {journal} {ArXiv e-prints}\ }
  (\bibinfo {year} {2013}{\natexlab{e}})},\ \Eprint
  {http://arxiv.org/abs/1303.5072} {arXiv:1303.5072 [astro-ph.CO]} \BibitemShut
  {NoStop}%
\bibitem [{\citenamefont {Aghanim}\ \emph
  {et~al.}(2013{\natexlab{b}})\citenamefont {Aghanim} \emph
  {et~al.}}]{Planck_LFI_beam}%
  \BibitemOpen
  \bibfield  {author} {\bibinfo {author} {\bibfnamefont {N.}~\bibnamefont
  {Aghanim}} \emph {et~al.} (\bibinfo {collaboration} {Planck Collaboration}),\
  }\href@noop {} {\bibfield  {journal} {\bibinfo  {journal} {ArXiv e-prints}\ }
  (\bibinfo {year} {2013}{\natexlab{b}})},\ \Eprint
  {http://arxiv.org/abs/1303.5065} {arXiv:1303.5065 [astro-ph.CO]} \BibitemShut
  {NoStop}%
\bibitem [{\citenamefont {Ade}\ \emph {et~al.}(2013{\natexlab{f}})\citenamefont
  {Ade} \emph {et~al.}}]{Planck_HFI_beam}%
  \BibitemOpen
  \bibfield  {author} {\bibinfo {author} {\bibfnamefont {P.}~\bibnamefont
  {Ade}} \emph {et~al.} (\bibinfo {collaboration} {Planck Collaboration}),\
  }\href@noop {} {\bibfield  {journal} {\bibinfo  {journal} {ArXiv e-prints}\ }
  (\bibinfo {year} {2013}{\natexlab{f}})},\ \Eprint
  {http://arxiv.org/abs/1303.5068} {arXiv:1303.5068 [astro-ph.IM]} \BibitemShut
  {NoStop}%
\bibitem [{\citenamefont {{Groeneboom}}\ and\ \citenamefont
  {{Eriksen}}(2009)}]{WMAP5_Eriksen_gstar1}%
  \BibitemOpen
  \bibfield  {author} {\bibinfo {author} {\bibfnamefont {N.~E.}\ \bibnamefont
  {{Groeneboom}}}\ and\ \bibinfo {author} {\bibfnamefont {H.~K.}\ \bibnamefont
  {{Eriksen}}},\ }\href {\doibase 10.1088/0004-637X/690/2/1807} {\bibfield
  {journal} {\bibinfo  {journal} {\apj}\ }\textbf {\bibinfo {volume} {690}},\
  \bibinfo {pages} {1807} (\bibinfo {year} {2009})},\ \bibinfo {note}
  {arXiv:0807.2242}\BibitemShut {NoStop}%
\bibitem [{\citenamefont {{Groeneboom}}\ \emph {et~al.}(2010)\citenamefont
  {{Groeneboom}}, \citenamefont {{Ackerman}}, \citenamefont {{Kathrine
  Wehus}},\ and\ \citenamefont {{Eriksen}}}]{WMAP5_Eriksen_gstar2}%
  \BibitemOpen
  \bibfield  {author} {\bibinfo {author} {\bibfnamefont {N.~E.}\ \bibnamefont
  {{Groeneboom}}}, \bibinfo {author} {\bibfnamefont {L.}~\bibnamefont
  {{Ackerman}}}, \bibinfo {author} {\bibfnamefont {I.}~\bibnamefont {{Kathrine
  Wehus}}}, \ and\ \bibinfo {author} {\bibfnamefont {H.~K.}\ \bibnamefont
  {{Eriksen}}},\ }\href {\doibase 10.1088/0004-637X/722/1/452} {\bibfield
  {journal} {\bibinfo  {journal} {\apj}\ }\textbf {\bibinfo {volume} {722}},\
  \bibinfo {pages} {452} (\bibinfo {year} {2010})},\ \Eprint
  {http://arxiv.org/abs/arXiv:0911.0150} {arXiv:0911.0150} \BibitemShut
  {NoStop}%
\bibitem [{\citenamefont {{Hanson}}\ \emph {et~al.}(2010)\citenamefont
  {{Hanson}}, \citenamefont {{Lewis}},\ and\ \citenamefont
  {{Challinor}}}]{gstar_beam}%
  \BibitemOpen
  \bibfield  {author} {\bibinfo {author} {\bibfnamefont {D.}~\bibnamefont
  {{Hanson}}}, \bibinfo {author} {\bibfnamefont {A.}~\bibnamefont {{Lewis}}}, \
  and\ \bibinfo {author} {\bibfnamefont {A.}~\bibnamefont {{Challinor}}},\
  }\href {\doibase 10.1103/PhysRevD.81.103003} {\bibfield  {journal} {\bibinfo
  {journal} {\prd}\ }\textbf {\bibinfo {volume} {81}},\ \bibinfo {pages}
  {103003} (\bibinfo {year} {2010})},\ \Eprint {http://arxiv.org/abs/1003.0198}
  {1003.0198} \BibitemShut {NoStop}%
\bibitem [{\citenamefont {{Bennett}}\ \emph {et~al.}(2013)\citenamefont
  {{Bennett}} \emph {et~al.}}]{WMAP9:results}%
  \BibitemOpen
  \bibfield  {author} {\bibinfo {author} {\bibfnamefont {C.~L.}\ \bibnamefont
  {{Bennett}}} \emph {et~al.},\ }\href {\doibase 10.1088/0067-0049/208/2/20}
  {\bibfield  {journal} {\bibinfo  {journal} {\apjs}\ }\textbf {\bibinfo
  {volume} {208}},\ \bibinfo {eid} {20} (\bibinfo {year} {2013})},\ \bibinfo
  {note} {arXiv:1212.5225}\BibitemShut {NoStop}%
\bibitem [{\citenamefont {Hill}\ \emph {et~al.}(2009)\citenamefont {Hill} \emph
  {et~al.}}]{Hill:2008}%
  \BibitemOpen
  \bibfield  {author} {\bibinfo {author} {\bibfnamefont {R.}~\bibnamefont
  {Hill}} \emph {et~al.} (\bibinfo {collaboration} {WMAP Collaboration}),\
  }\href {\doibase 10.1088/0067-0049/180/2/246} {\bibfield  {journal} {\bibinfo
   {journal} {Astrophys.J.Suppl.}\ }\textbf {\bibinfo {volume} {180}},\
  \bibinfo {pages} {246} (\bibinfo {year} {2009})},\ \Eprint
  {http://arxiv.org/abs/0803.0570} {arXiv:0803.0570 [astro-ph]} \BibitemShut
  {NoStop}%
\bibitem [{\citenamefont {{Mitra}}\ \emph {et~al.}(2011)\citenamefont
  {{Mitra}}, \citenamefont {{Rocha}}, \citenamefont {{G{\'o}rski}},
  \citenamefont {{Huffenberger}}, \citenamefont {{Eriksen}}, \citenamefont
  {{Ashdown}},\ and\ \citenamefont {{Lawrence}}}]{effConv}%
  \BibitemOpen
  \bibfield  {author} {\bibinfo {author} {\bibfnamefont {S.}~\bibnamefont
  {{Mitra}}}, \bibinfo {author} {\bibfnamefont {G.}~\bibnamefont {{Rocha}}},
  \bibinfo {author} {\bibfnamefont {K.~M.}\ \bibnamefont {{G{\'o}rski}}},
  \bibinfo {author} {\bibfnamefont {K.~M.}\ \bibnamefont {{Huffenberger}}},
  \bibinfo {author} {\bibfnamefont {H.~K.}\ \bibnamefont {{Eriksen}}}, \bibinfo
  {author} {\bibfnamefont {M.~A.~J.}\ \bibnamefont {{Ashdown}}}, \ and\
  \bibinfo {author} {\bibfnamefont {C.~R.}\ \bibnamefont {{Lawrence}}},\ }\href
  {\doibase 10.1088/0067-0049/193/1/5} {\bibfield  {journal} {\bibinfo
  {journal} {\apjs}\ }\textbf {\bibinfo {volume} {193}},\ \bibinfo {eid} {5}
  (\bibinfo {year} {2011})},\ \Eprint {http://arxiv.org/abs/1005.1929}
  {arXiv:1005.1929 [astro-ph.CO]} \BibitemShut {NoStop}%
\bibitem [{\citenamefont {Gorski}\ \emph {et~al.}(2005)\citenamefont {Gorski},
  \citenamefont {Hivon}, \citenamefont {Banday}, \citenamefont {Wandelt},
  \citenamefont {Hansen}, \citenamefont {Reinecke},\ and\ \citenamefont
  {Bartelman}}]{HEALPix:framework}%
  \BibitemOpen
  \bibfield  {author} {\bibinfo {author} {\bibfnamefont {K.~M.}\ \bibnamefont
  {Gorski}}, \bibinfo {author} {\bibfnamefont {E.}~\bibnamefont {Hivon}},
  \bibinfo {author} {\bibfnamefont {A.~J.}\ \bibnamefont {Banday}}, \bibinfo
  {author} {\bibfnamefont {B.~D.}\ \bibnamefont {Wandelt}}, \bibinfo {author}
  {\bibfnamefont {F.~K.}\ \bibnamefont {Hansen}}, \bibinfo {author}
  {\bibfnamefont {M.}~\bibnamefont {Reinecke}}, \ and\ \bibinfo {author}
  {\bibfnamefont {M.}~\bibnamefont {Bartelman}},\ }\href@noop {} {\bibfield
  {journal} {\bibinfo  {journal} {\apj}\ }\textbf {\bibinfo {volume} {622}},\
  \bibinfo {pages} {759} (\bibinfo {year} {2005})}\BibitemShut {NoStop}%
\bibitem [{\citenamefont {Lewis}\ \emph {et~al.}(2000)\citenamefont {Lewis},
  \citenamefont {Challinor},\ and\ \citenamefont {Lasenby}}]{CAMB}%
  \BibitemOpen
  \bibfield  {author} {\bibinfo {author} {\bibfnamefont {A.}~\bibnamefont
  {Lewis}}, \bibinfo {author} {\bibfnamefont {A.}~\bibnamefont {Challinor}}, \
  and\ \bibinfo {author} {\bibfnamefont {A.}~\bibnamefont {Lasenby}},\
  }\href@noop {} {\bibfield  {journal} {\bibinfo  {journal} {\apj}\ }\textbf
  {\bibinfo {volume} {538}},\ \bibinfo {pages} {473} (\bibinfo {year}
  {2000})},\ \Eprint {http://arxiv.org/abs/http://camb.info/}
  {http://camb.info/} \BibitemShut {NoStop}%
\end{thebibliography}%
\end{document}